\documentclass{ws-procs9x6}
\def\simle{
    \mathrel{\rlap{\raise 0.511ex
        \hbox{$<$}}{\lower 0.511ex \hbox{$\sim$}}}}
\def\simge{
    \mathrel{\rlap{\raise 0.511ex
        \hbox{$>$}}{\lower 0.511ex \hbox{$\sim$}}}}
\usepackage{graphicx}

\begin{document}

\title{Lattice calculation of the lowest order hadronic contribution to the
muon anomalous magnetic moment: an update with Kogut-Susskind fermions
\footnote{\uppercase{I} thank the \uppercase{US DOE, RIKEN},
and \uppercase{NERSC} for providing resources to complete this work.}
}

\author{T. Blum}

\address{RIKEN BNL Research Center\\ Brookhaven National Laboratory\\
        Upton NY 11973-5000, USA\\ E-mail: tblum@bnl.gov}

\maketitle

\abstracts{I present a preliminary calculation of the hadronic vacuum
polarization for 2+1 flavors of improved Kogut-Susskind quarks by utilizing a
set of gauge configurations recently generated by the MILC collaboration. The
polarization function $\Pi(q^2)$ is then used to calculate the lowest order
(in $\alpha_{QED}$) hadronic contribution to the muon anomalous magnetic
moment.}

\section{Introduction}

The anomalous magnetic moment of the muon, $a_{\mu}$, is now known to
fantastic precision, both experimentally\cite{Bennett:2002jb} and
theoretically\cite{Lepton-Moments}. This situation naturally provides an
interesting test of the Standard Model.  The largest uncertainty in the
calculation of $a_{\mu}$ is associated with hadronic contributions; the lowest
order in $\alpha$ contribution arises from the hadronic vacuum polarization of
the photon.  Using the analytic structure of the vacuum polarization and the
optical theorem, this contribution is estimated from the experimentally
measured total cross section of $e^+e^-$ annihilation to
hardrons\cite{Davier:2002dy,Hagiwara:2002ma}.  Isospin symmetry relates the
$e^+e^-$ cross-section to the branching ratio of $\tau$ decay to hadrons which
can also be used to calculate the hadronic
contributions\cite{Davier:2002dy}. However, a purely theoretical, first
principles treatment has been missing. Given the importance of the muon g-2
experiment, a completely independent theoretical calculation is desirable.

Recently, the framework to calculate the hadronic piece of the anomalous
moment, $a^{\rm had}_\mu$, in an entirely theoretical way from first
principles using lattice QCD was given\cite{Blum:2002ii}. Encouraging
quenched results were obtained using domain wall
fermions\cite{Blum:2002ii} and improved Wilson fermions\cite{Rakow}.

In a nut-shell, the first principles calculation is performed entirely in
Euclidean space-time, so that the hadronic vacuum polarization computed on the
lattice can be inserted directly into the one-loop vertex function for the
muon. To obtain the physical result, at the end one continues back to Minkowski
space-time so the external muon is on-shell.

The aim of the current study is to determine how accurate the lattice
calculation can be in the near future. Toward that goal, I present preliminary
results for the hadronic vacuum polarization calculated on a set of 2+1 flavor
improved Kogut-Susskind fermion lattices generated by the MILC collaboration
(see Table~\ref{table:1}).

\section{The vacuum polarization}

The vacuum polarization tensor is defined as the Fourier transform of the
two-point correlation function of the electromagnetic current, $J^\mu$, 
\begin{eqnarray}
\Pi^{\mu\nu}(q^2) &=& \int d^4x\, e^{iq(x-y)}J^\mu(x)\,J^\nu(y),\\ &=& (q^\mu
q^\nu - \delta^{\mu\nu})\Pi(q^2),\\ J^\mu &=&
\frac{2}{3}\bar{u}\,\gamma_\mu\,u-
\frac{1}{3}\bar{d}\,\gamma_\mu\,d-
\frac{1}{3}\bar{s}\,\gamma_\mu\,s.
\label{eq:pimunu}
\end{eqnarray}
To satisfy the Ward-Takahashi identity, 
\begin{eqnarray}
q^\mu\,\Pi^{\mu\nu}=0,
\end{eqnarray} 
the exactly conserved lattice vector current is used which makes the
extraction of the polarization function from Eq.~(\ref{eq:pimunu})
straightforward.  In general, the lattice conserved current is point-split; it
depends on the fields at a point $x$ and its neighbors:
\begin{eqnarray}
J^\mu(x) &=&
\frac{c_0}{2}\left(\bar\psi(x+\hat\mu)\,U^\dagger_\mu(x)\,\gamma_\mu\psi(x)+
\bar\psi(x)\,U_\mu(x)\,\gamma_\mu\psi(x+\hat\mu)\right)
\end{eqnarray}
 (for a single flavor) and satisfies 
$\Delta^\mu J^\mu= 0.$
The gauge field $U_\mu(x)$ makes the current gauge-invariant, and $c_0$
depends on the lattice action ($c_0=1$ for ordinary Kogut-Susskind fermions).
With this form for the lattice current, and after subtracting the contact
terms that arise because of its point-split
form\cite{Gockeler:2000kj,Blum:2002ii}, the lattice polarization tensor is
given by Eq. 2 with $q^\mu=2\sin{(\pi\, n_\mu /L_\mu)}$,
$n_\mu=0,1,2,...,L_\mu-1$.  Adding to the lattice action the three-hop Naik
term\cite{Naik:1986bn} to improve the discretization of the derivative in the
Dirac operator generates an additional divergence. In coordinate space
\begin{eqnarray}
J^{3\mu}(x) &=& 
\frac{c_1}{2}\left(
\bar\psi(x+3\hat\mu)V_\mu^\dagger(x)\gamma_\mu\psi(x)+
\bar\psi(x)V_\mu(x)\gamma_\mu\psi(x+3\hat\mu)
\right),\\
V_\mu(x) &=& U_\mu(x)U_\mu(x+\hat\mu)U_\mu(x+2\hat\mu),
\end{eqnarray}
with $c_0=9/8$, $c_1=-1/24$ for Naik fermions,
and which satisfies $\Delta^\mu J^\mu+ \Delta^{3\mu} J^{3\mu}= 0$.
The backward difference operators are defined as 
\begin{eqnarray}
\Delta^\mu f(x) & = & f(x) - f(x-\hat\mu),\\
\Delta^{3\mu} f(x) & = & f(x) - f(x-3\hat\mu).
\end{eqnarray}
Thus the simple form of the Ward-Takahashi identity, and hence $\Pi^{\mu\nu}$,
no longer holds. To avoid this complication, in the following the Naik term is
dropped when computing valence quark propagators even though it appears in the
$a^2$-tad action\cite{ASQTAD} used to generate the 2+1 flavor
configurations. Note that the discretization errors are then ${O}(a^2)$, not
${O}(g^2\,a^2)$ as for the full $a^2$-tad action. This modification should not
significantly affect the small $q^2$, or long-distance, behavior of
$\Pi(q^2)$.

\begin{table}
\tbl{MILC 2+1 flavor lattices generated with the $a^2$-tad
  action\protect\cite{ASQTAD}. $a$ is the lattice
  spacing\protect\cite{Doug}. $m_l$ and $m_s$ denote light and strange quark
  masses, respectively. In all cases the strange quark mass corresponds
  roughly to its physical value. $m_v$ and \# configs are the valence quark
  mass and number of configurations used in the calculation of $\Pi(q^2)$.
\label{table:1}}
{\begin{tabular}{@{}cccccc@{}}
\hline
$a$ (fm) & size   & $m_l$  & $m_s$ & $m_{val}$ & \# configs\cr
\hline
0.121(3) & $20^3\times 64$ & 0.01   & 0.05  & 0.05      & 57\cr
0.121(3) & $20^3\times 64$ & 0.01   & 0.05  & 0.01      & 439\cr
0.120(3) & $24^3\times 64$ & 0.005  & 0.05  & 0.005     & 143\cr
0.086(2) & $28^3\times 96$ & 0.0062 & 0.031 & 0.031     & 41\cr
0.086(2) & $28^3\times 96$ & 0.0062 & 0.031 & 0.0062    & 248\cr
0.085(2) & $28^3\times 96$ & \multicolumn{2}{l}{quenched} & 0.031   & 29\cr
0.085(2) & $28^3\times 96$ & \multicolumn{2}{l}{quenched} & 0.0062  & 31\cr
\hline
\hline
\end{tabular}\\[2pt]}
\end{table}

\section{RESULTS and DISCUSSION}

\subsection{vacuum polarization}

\begin{figure}[ht]
\vskip .25in
\centerline{\epsfxsize=4.in
\includegraphics[width=\epsfxsize]{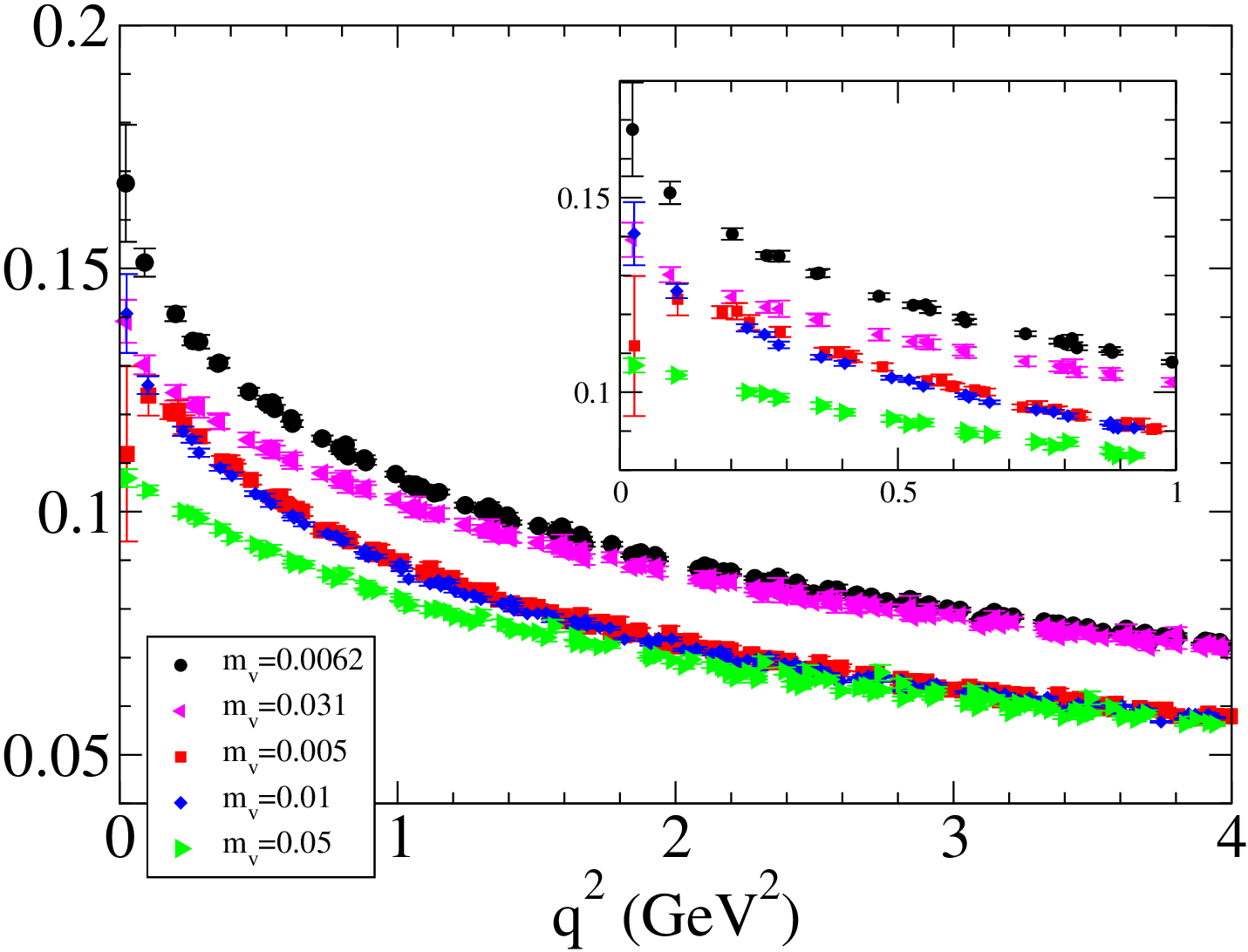}}
\caption{$\Pi(q^2)$ computed on 2+1 flavor lattices. The two upper curves
  correspond to the finer lattice spacing. Errors shown are statistical only.
  \label{fig:1}}
\end{figure}

In Figure~\ref{fig:1}, I show $\Pi(q^2)$ calculated on the improved
Kogut-Susskind lattices\footnote{Some of these results were obtained after the
meeting.}.  All results at a given lattice spacing approach a common value in
the large $q^2$ limit, as they should. $\Pi(q^2)$ is logarithmically
divergent in the lattice spacing, only the running with $q^2$ being
physical. Therefore, the results for different lattice spacings differ by a
constant plus terms of order $a^2$.  Such a shift is clearly visible between
the two sets of data plotted in Figure~\ref{fig:1}. For the coarser lattice
spacing ($a\approx 0.121$ fm), $\Pi(q^2)$ was calculated for three valence
quark masses, $m_v=0.005$, 0.01, and 0.05, or $m_v\approx 0.1\,m_s$,
$0.2\,m_s$, and $m_s$ respectively. For the finer lattice ($a\approx 0.086$
fm) we have $m_v=0.0062$ and 0.031, or $m_v\approx 0.2\,m_s$ and $m_s$. There
is a significant dependence on the valence quark mass. As $q^2\to 0$,
$\Pi(q^2)$ rises much more steeply as $m_v\to 0$. However, for the larger
lattice spacing this behavior appears to weaken sharply between $m_v=0.01$ and
0.005. For $m_v$= 0.005 $\Pi(q^2)$ actually decreases at the lowest value of
$q^2$ which is probably indicative of low statistics.

\begin{figure}[ht]
\vskip .25in
\centerline{\epsfxsize=4.in
\includegraphics[width=\epsfxsize]{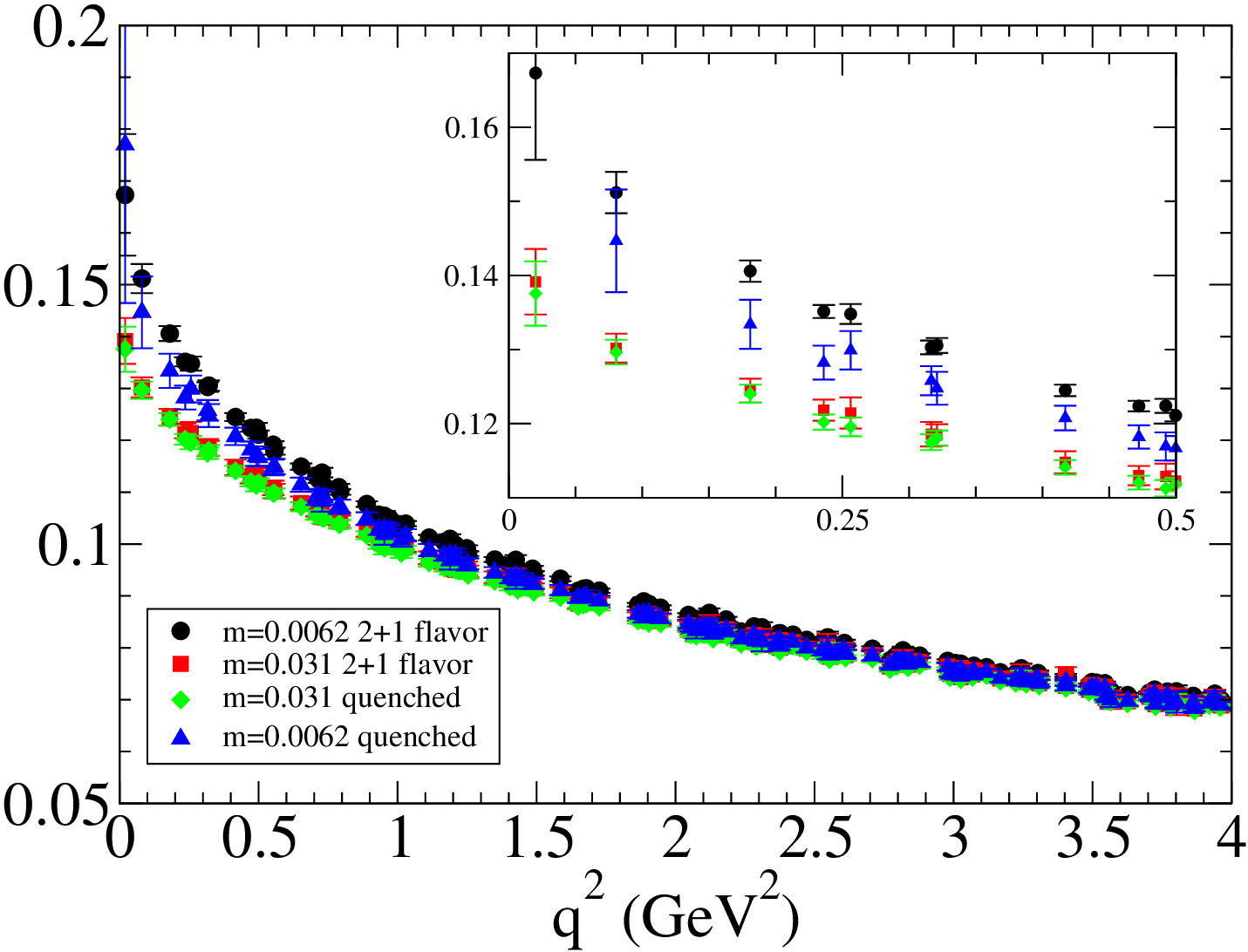}}
\caption{$\Pi(q^2)$. 2+1 flavor and quenched
         lattices at $a\approx0.086$ fm.
         \label{fig:2}}
\end{figure}

In Figure~\ref{fig:2}, I compare $\Pi(q^2)$ computed on dynamical and quenched
lattices at $a\approx 0.086$ fm ($a^{-1}\approx 2.239$ GeV).  Discernible
effects appear as $q^2\to 0$ for $m_v=0.0062$, the lightest quark mass for
this lattice spacing. More statistics on the quenched lattice are needed to
quantify the effect. For $m_v=0.031$ there is no apparent effect of
unquenching. These results may indicate that the two-pion threshold is not
lower than the mass of the vector particle.  The two-pion state must have one
unit of orbital angular momentum since the photon has $J=1$, so the $\pi$'s
can not be at rest. The threshold for two non-interacting pions is $2\,E_\pi=
2\,\sqrt{m_\pi^2 + (2\pi/L)^2}$, or $\approx 0.54$ and 0.61 for $m_v=0.0062$
and 0.005, respectively.  On the other hand, $m_\rho\approx 0.39$ and 0.53 in
these cases\cite{Doug}. Note that since $\Pi(q^2)$ is computed in Euclidean
space, it should be a smooth function of $q^2$, even as such thresholds and
resonances are crossed (in Minkowski space)\cite{Eduardo}.

\begin{figure}[ht]
\vskip .25in
\centerline{\epsfxsize=3.8in
\includegraphics[width=\epsfxsize]{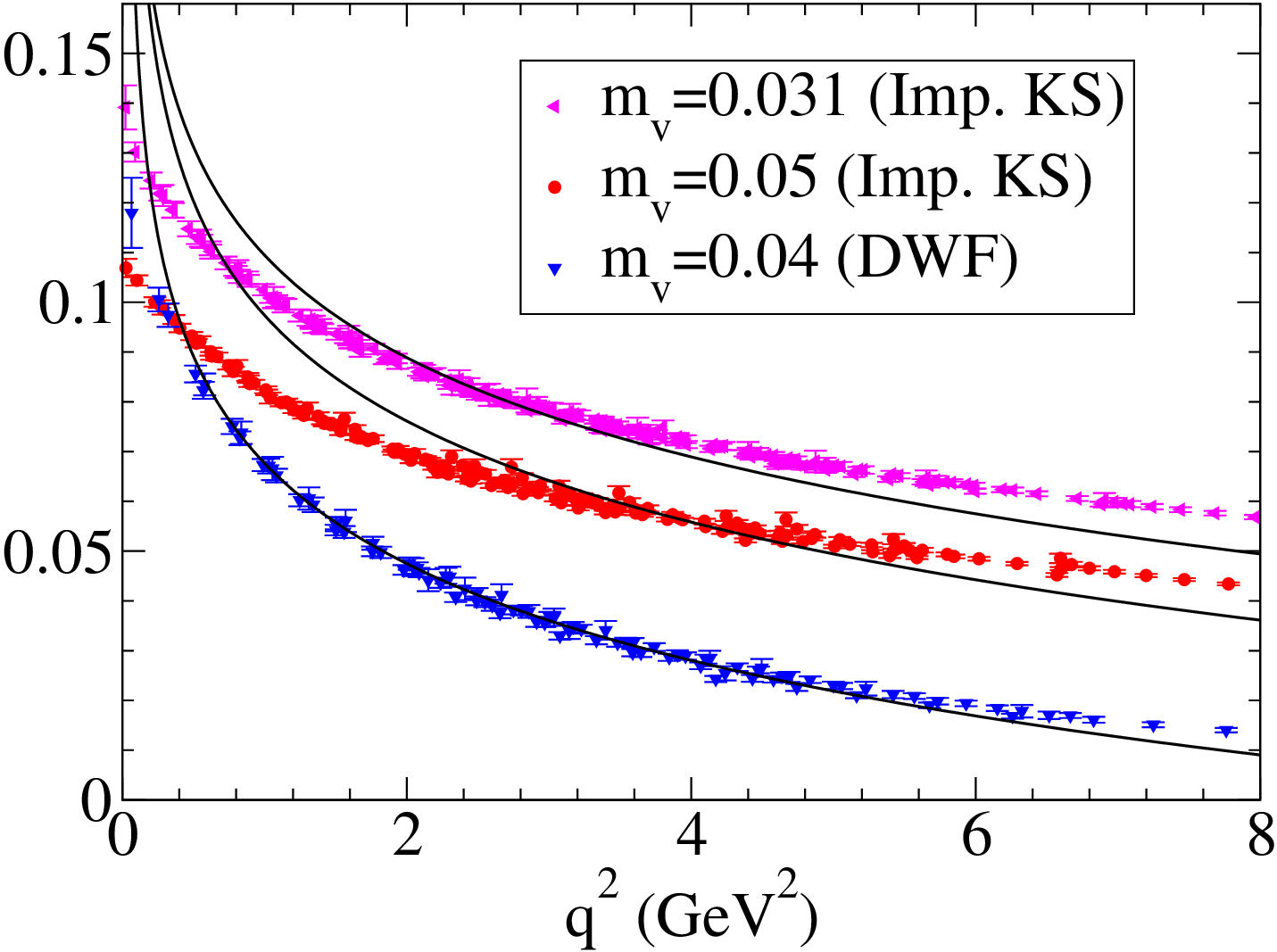}}
\caption{$\Pi(q^2)$. 2+1 flavor lattices. Comparison to continuum perturbation
         theory\protect\cite{Chetyrkin:1996cf}. $m_v=0.05$ ($a=0.121$ fm),
         0.031 ($a=0.086$ fm), and 0.04 (quenched domain wall fermion results
         from \protect\cite{Blum:2002ii}, $a=0.15$ fm).
         \label{fig:3}}
\end{figure}

\begin{figure}[ht]
\vskip .25in
\centerline{\epsfxsize=4.in
\includegraphics[width=\epsfxsize]{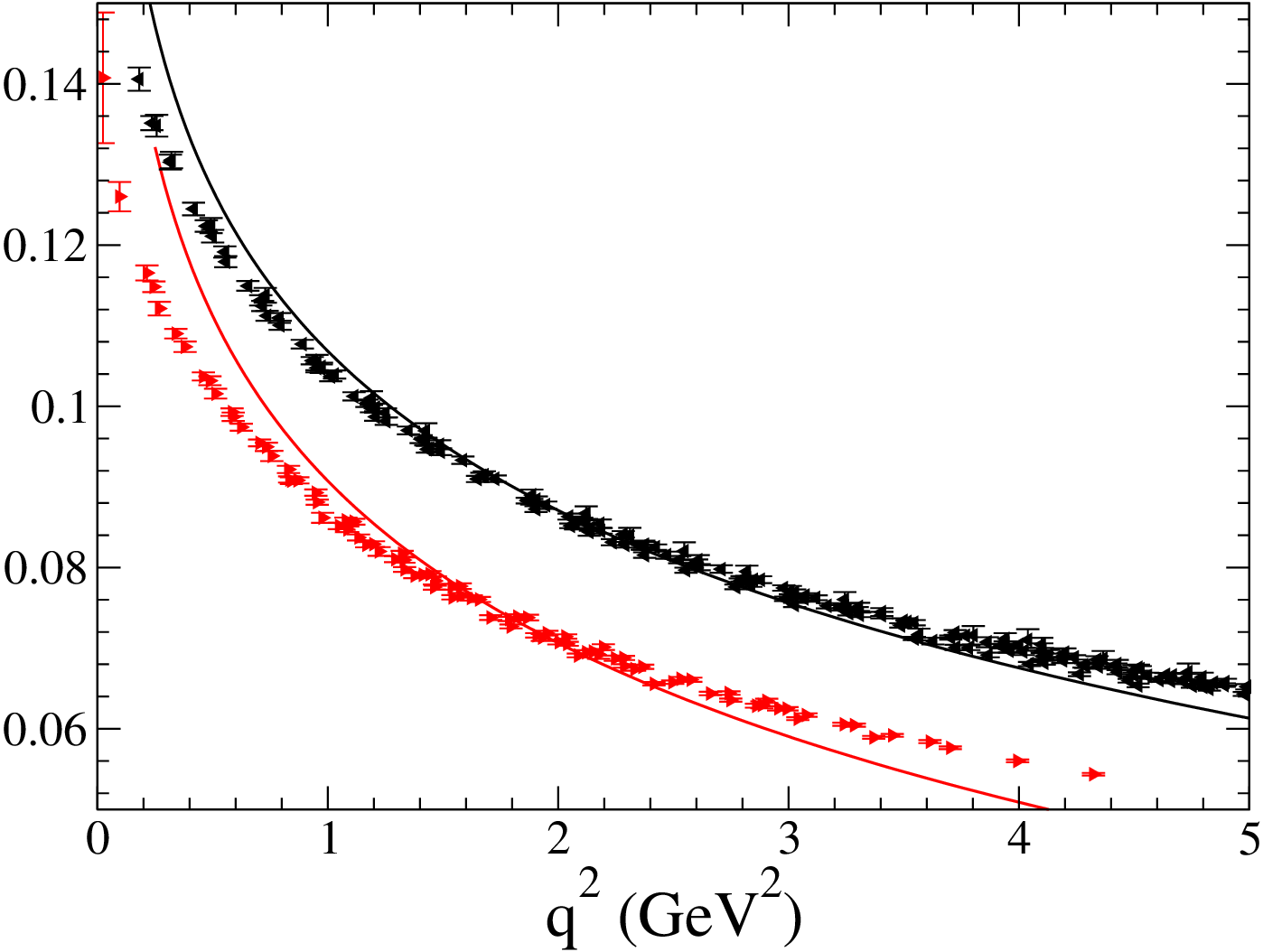}}
\caption{$\Pi(q^2)$. Same as Figure~\protect\ref{fig:3}, but for $m_v=0.01$
  (lower) and 0.0062 (upper). The lattice (symbols) and perturbation theory
  (lines) results are arbitrarily matched at $q^2=2$ GeV$^2$ for
  comparison. Agreement with perturbation theory is better for the smaller
  lattice spacing, $a=0.086$ fm (upper points).
  \label{fig:4}}
\end{figure}

Comparison of $\Pi(q^2)$ with continuum perturbation
theory\cite{Chetyrkin:1996cf} as shown in Figure~\ref{fig:3} indicates that
the improved Kogut-Susskind results may suffer significant lattice artifacts
(contrast with the quenched domain wall fermion results at roughly the same
$m_v$ but larger lattice spacing). Results are shifted by hand to account for
the ln$(a)$ term; there is no choice for this shift which yields good
agreement over a large range of $q^2$ with continuum perturbation theory,
unlike the case for domain wall fermions which compares quite well
The
agreement does improve as $m_v\to 0$ (Figure~\ref{fig:4}), in particular for
the smaller lattice spacing. In Figure~\ref{fig:4} I have chosen,
arbitrarily, to match the lattice data and the continuum perturbation theory
at $q^2=2$ GeV$^2$. This behavior may be indicative of an $a^2\,m\,q$
error. The perturbation theory results shown in Figures~\ref{fig:3}
and~\ref{fig:4} are given in the $\overline{MS}$ scheme
\footnote{Of course, the physical piece of $\Pi(q^2)$ is scheme and scale
independent.}
, so $m_v$ should also be given in this scheme. Except for the
domain wall fermion results, this has not been done, but note that the
quark mass dependence of the perturbation theory result is very mild for small
mass. 

\begin{figure}[ht]
\vskip .3in
\centerline{\epsfxsize=4.in
\includegraphics[width=\epsfxsize]{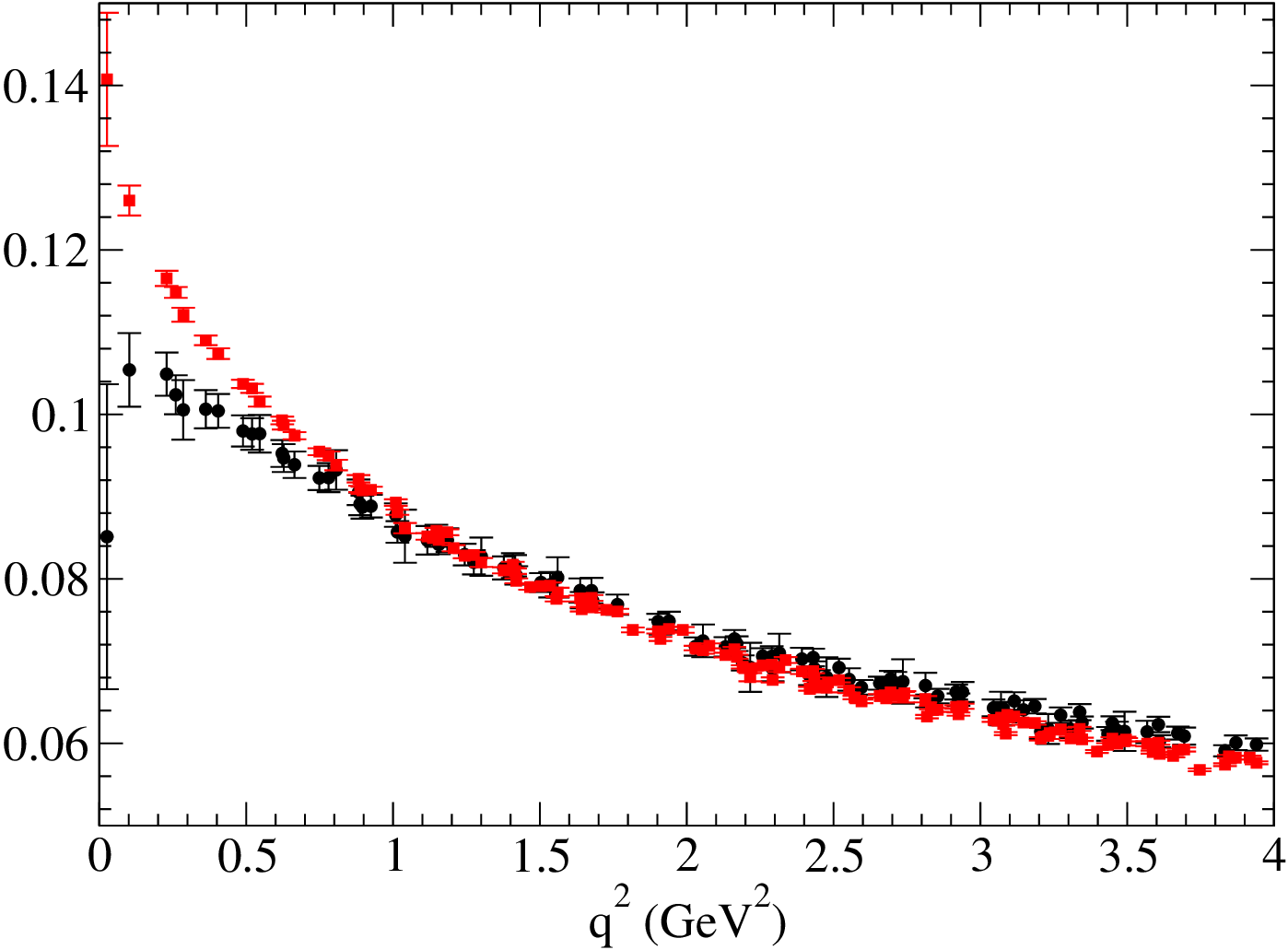}}
\caption{$\Pi(q^2)$. Comparison of fat-link improved ($a^2$-tad minus the Naik
  term) and ordinary Kogut-Susskind fermions. The smaller
  slope as $q^2\to0$ is indicative of lattice spacing artifacts that
  arise from flavor symmetry breaking.\label{fig:5}} 
\end{figure}

The valence Kogut-Susskind fermions contribute like four flavors of
continuum fermions to $\Pi(q^2)$ which therefore has to be scaled by
1/4. In the continuum limit, $a\to 0$, these four flavors are
degenerate, so the scaling is exact. For $a\neq0$ this scaling is not
exact and leads to lattice spacing errors like the ones discussed
above. It is just such flavor-symmetry breaking artifacts that are
supposed to be suppressed by the $a^2$-tad action.  Figure~\ref{fig:5} shows
that, indeed, the fat-link Kogut-Susskind fermions differ
significantly from ordinary Kogut-Susskind fermions; the
slope increases both at low and high values of $q^2$.

\begin{figure}[ht]
\vskip .25in
\centerline{\epsfxsize=4.in
\includegraphics[width=\epsfxsize]{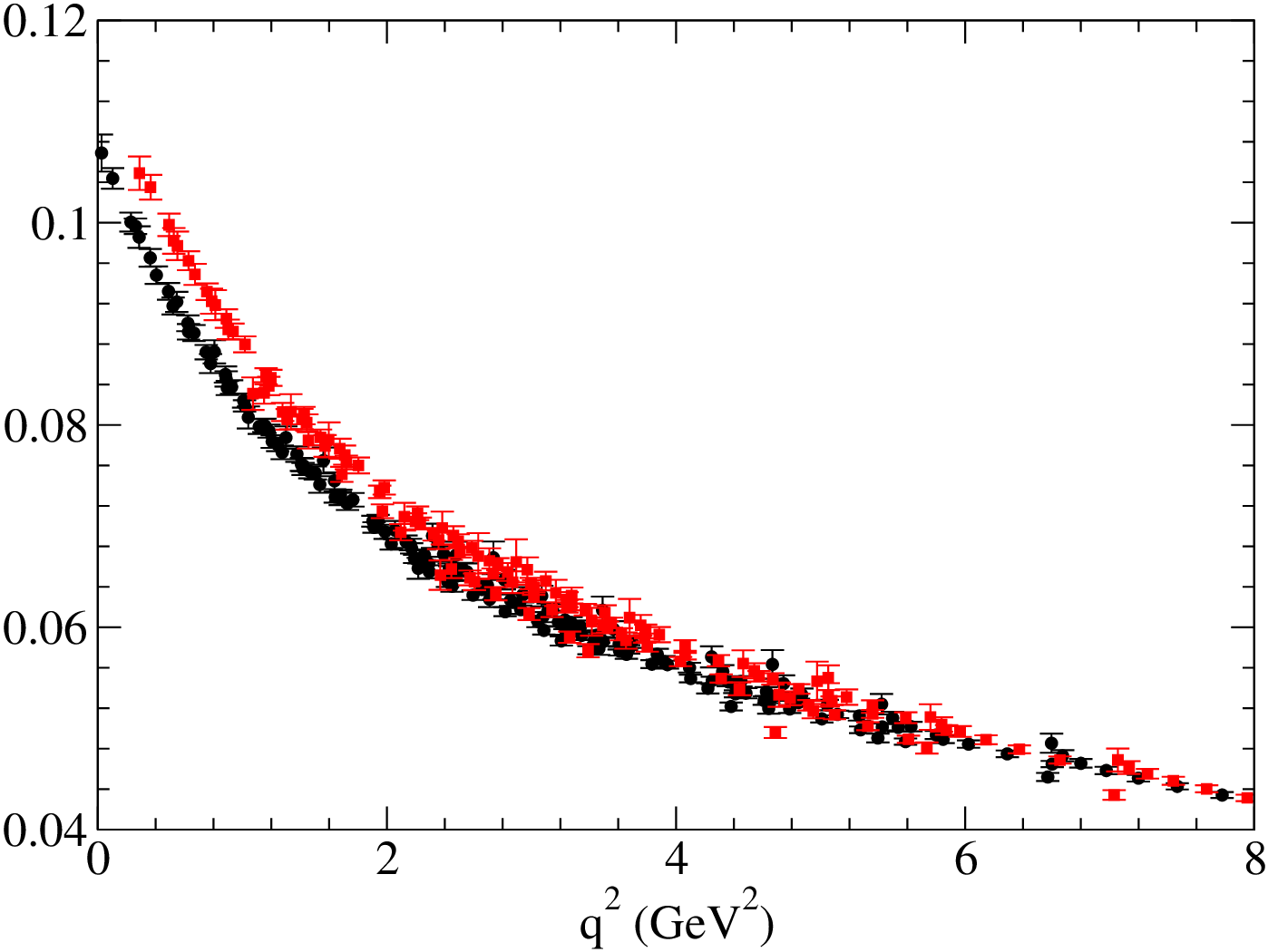}}
\caption{Effect of the Naik term (upper points) on $\Pi(q^2)$.
$m_v=0.05$, $m_l=0.01$, and $m_s=0.05$. \label{fig:6}}
\end{figure}

As mentioned above, omission of the Naik term should not significantly alter
the low $q^2$ behavior of $\Pi(q^2)$, even though the errors are now order
$a^2$, not $g^2\, a^2$.  This is because the Naik term improves the derivative
in the Dirac operator, but does not correct flavor symmetry.  The reason for
omitting the Naik term is that the simple ansatz, Eq. 2, no longer holds.
Still, this ansatz should be a good estimate even for Naik fermions for small
$q$.  In Figure~\ref{fig:6}, $\Pi(q^2)$ ($m_v=0.01$, $a=0.121$ fm) is shown
with and without the Naik term. For small $q^2$ the results appear similar,
apart from a constant shift.  There is also a hint that the Naik term makes
$\Pi(q^2)$ steeper at larger values of $q^2$ which may improve the agreement
with perturbation theory. However because only an approximate ansatz for
$\Pi^{\mu\nu}$ is used in this case, this can only be taken as an indication.
\vspace{-.25in}

\subsection{The anomalous magnetic moment}
\begin{figure}[h]
\vskip .25in
\centerline{\epsfxsize=4.in
\includegraphics[width=\epsfxsize]{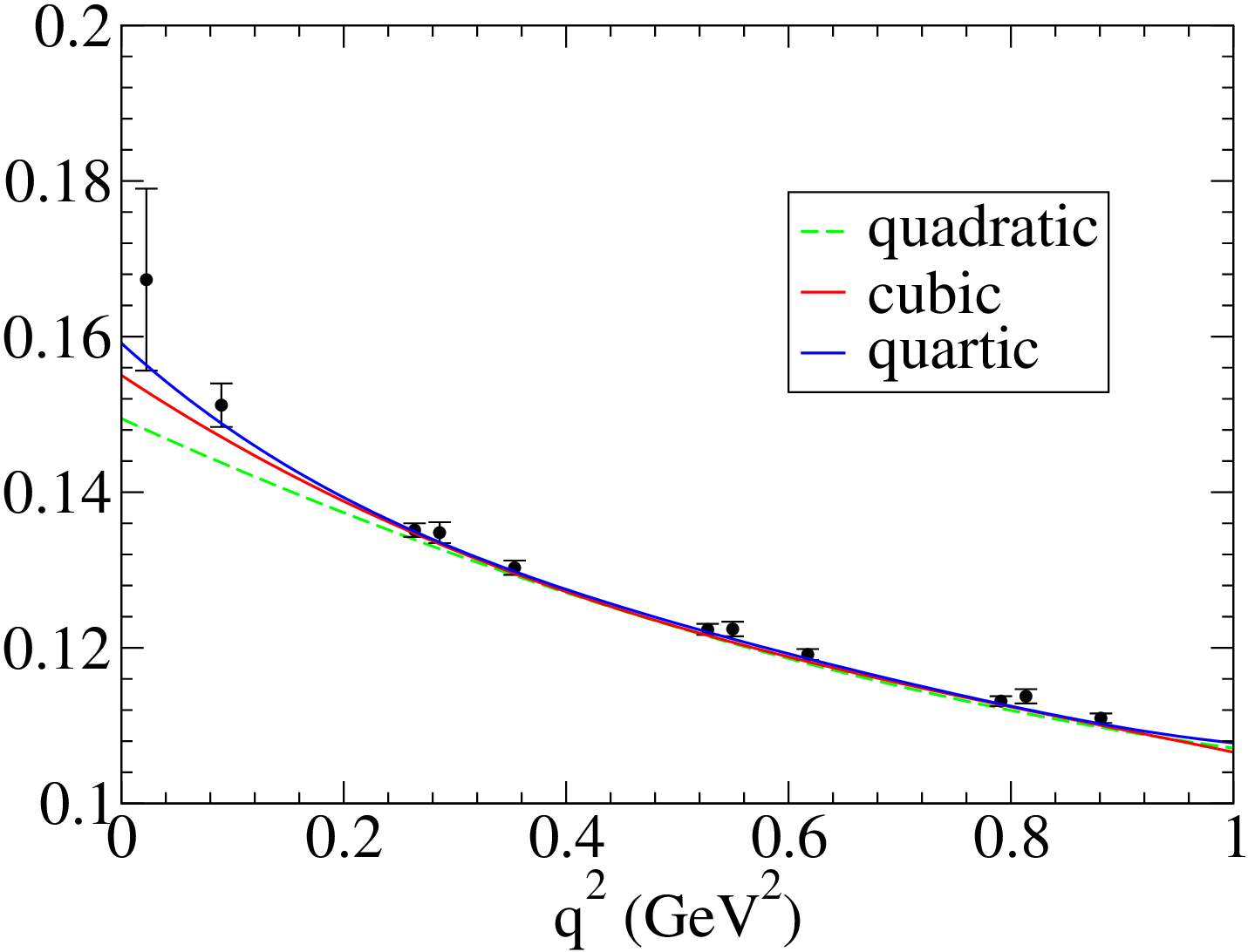}}
\caption{Covariant fits to $\Pi(q^2)$. 2+1 flavor lattice,
$m_v=0.0062$.\label{fig:7}}
\end{figure}

The method for calculating $a^{\rm had}_\mu$ from the vacuum polarization is
given in \cite{Blum:2002ii}. The central idea is to carry out the entire
calculation in Euclidean space so that $\Pi(q^2)$ calculated on the lattice
up to some momentum, $q^2_{\rm cut}$,
can be directly inserted into the one-loop (QED) vertex of the muon which
describes its interaction with an external magnetic field.  A key feature is
that the integral is dominated by the low $q^2$ region.  If the lattice
calculation is accurate enough, {\it i.e.} statistical errors are under good
control, $q^2$ is low enough, {\it etc.}, then no additional theoretical input
like a fit ansatz is necessary, and any faithful representation of the
numerical data will suffice to calculate $a^{\rm had}_\mu$.  I have chosen a
simple polynomial to fit the lattice data,
\begin{eqnarray}
a^{\rm had}_\mu &= & a_0 + a_1\,x + a_2\, x^2 + a_3\,x^3 + a_4\,x^4,\\\nonumber
x &=& q^2.
\end{eqnarray}
The results for $m_v=0.0062$, $a=0.086$ fm are shown in Figure~\ref{fig:7}.
Values of $\Pi(q^2)$ at different $q^2$ are highly correlated, so 
I have fit only those results with $q^2\simle 1$ GeV$^2$ in order to obtain a
reasonable covariance matrix, and consequently, fits with acceptable $\chi^2$.
The fits shown in Figure~\ref{fig:7} are representative of all the data; they
tend to under-predict the data as $q^2\to0$. The discrepancy decreases as
more terms are added to the fit function. As only its running with $q^2$ is
physical, $\Pi(q^2)$ is renormalized by subtracting $\Pi(0)$, so the smaller
the slope of $\Pi(q^2\to0)$, the smaller is $a^{\rm had}_\mu$.

Table 2 summarizes the preliminary values for $a^{\rm had}_\mu$ computed using
improved Kogut-Susskind fermions on 2+1 flavor gauge configurations. The
lattice results are used up to $q^2_{\rm cut}=1$ GeV$^2$, and perturbation
theory from there to $\infty$ to complete the integral. The matching with
perturbation theory is not as good as for domain wall fermions, but since the
perturbative contribution is already quite small at 1 GeV$^2$ this will not
matter.  For comparison, quenched results are also tabulated, including the
domain wall fermion ones from \cite{Blum:2002ii}.  The dependence on the
polynomial degree of the fit function, seen in the fit results just described,
shows up in $a^{\rm had}_\mu$ due to the sensitivity of $a^{\rm had}_\mu$ to
the low $q^2$ region .  Compared to the quenched domain wall fermion value,
the improved Kogut-Susskind values are low, except the $m_v=0.0062$
point. This is probably a consequence of the lattice spacing errors in the
latter. For domain wall fermions, there is no indication of large scaling
violations\cite{Blum:2002ii}, though finite volume effects prevent ruling out
this possibility. Calculating the quark mass dependence with domain wall
fermions would also improve the comparison with the results presented here.

Using the quartic fit value, for 2+1 flavors, 
$m_{u,d}=0.0062\approx 1/5\, m_s$,
$m_s=0.031$,
$a=0.086$,
\begin{eqnarray*}
a^{\rm had}_{\mu,2+1}\,\times\,10^{10} &= &
\left(\left(\frac{2}{3}\right)^2+
\left(\frac{1}{3}\right)^2\right)\times903(117)+ 
\left(\frac{1}{3}\right)^2 \times393(72)\\\nonumber &=& 545(65).
\end{eqnarray*}
This should be taken only as a preliminary estimate.  The fits to $\Pi(q^2)$
are not very stable; a better fit method should not under-predict the low
$q^2$ region and may reduce the statistical errors.  Concentrating on the
light quark masses 0.0062 and 0.01 in Table 2, there is a significant increase
in $a^{\rm had}_\mu$ at $a= 0.086$ fm over the $a=0.121$ fm value; linear
extrapolation to $a^2=0$ increases $a^{\rm had}_\mu$ by $100\times 10^{-10}$.
Finally, the disconnected part of the vacuum polarization has not been
calculated. This piece is color and electric charge suppressed but could still
contribute to $a^{\rm had}_\mu$.  For comparison, the value computed from the
$e^+e^-$ cross-section is about $700\times 10^{-10}$
\cite{Davier:2003pw,Jegerlehner:2003qp}.
\begin{table}[h]
\tbl{$a_\mu^{\rm had}\times 10^{10}$ for a single quark flavor (entries
  correspond to Table 1). The eighth entry is for quenched domain wall
  fermions\protect\cite{Blum:2002ii}. Values obtained from covariant
  polynomial fits to $\Pi(q^2)$. The instability in some cases stems from the
  under-estimation of the magnitude of the slope of $\Pi(q^2)$ as $q^2\to 0$
  (see Figure~\ref{fig:7}). In all fits $q^2 < 1$ GeV$^2$.\label{table:2}}
  {\begin{tabular}{@{}lllll@{}} \hline $m_v$ & $m_{l}/m_s$ & quadratic & cubic
  & quartic \cr \hline 0.05 & 0.01/0.05 & 252(5)$^*$ & 310(11)$^*$ &
  252(56)$^*$\cr 0.01 & 0.01/0.05 & 433(10)& 543(30)& 610(84)\cr 0.005 &
  0.01/0.05 & 542(36)& 671(158)& 267(411)\cr 0.031 & 0.01/0.05 & 249(8)$^*$&
  383(24) & 393(72)\cr 0.0062 & 0.01/0.05 & 484(15)$^*$& 670(42) & 903(117)\cr
  0.031 & quenched & 277(7)& 315(18) & 349(80)\cr 0.0062 & quenched &
  393(34)$^*$& 274(98)$^*$& 297(203)$^*$\cr 0.04 & quenched & 655(43)& 772(52)
  & 840(299)\cr \hline \hline
\end{tabular}\\[2pt]}
$^*$ $\Pi(q^2)$ fit has poor $\chi^2$
\end{table}

\section{Summary}

Using gluon configurations generated with the $a^2$-tad lattice action, the
hadronic vacuum polarization was calculated for 2+1 flavor QCD.  The
contribution to the muon's anomalous magnetic moment was then obtained.  The
results are similar to quenched ones calculated previously, so more work,
mostly to reduce the light quark masses, must be done. It was found that the
improved Kogut-Susskind fermions still have significant lattice spacing
errors. While domain wall fermions appear to exhibit better scaling,
they are considerably more expensive than Kogut-Susskind fermions, especially
when considering the small quark mass and large volume limits.

It appears that to compete with the accuracy of the usual dispersive method,
which is quoted at about the one percent level, several improvements are still
needed: (1) better statistics in the low $q^2$ regime, (2) a more
accurate fit method, (3) smaller
quark masses so the vector particles are truly unstable and have their
physical widths, (4) a smaller lattice spacing if Kogut-Susskind fermions are
used.

I am encouraged by on-going MILC Collaboration simulations at $a= 0.086$ fm,
$m_l=0.0031$, $m_s=0.031$ on $40^3\times96$ lattices. This light quark mass is
two times smaller than the quark mass used here and approximately 10 times
lighter than the strange quark mass. Given the good scaling behavior of domain
wall fermions, it would also be interesting to compute $\Pi(q^2)$ on 2 flavor
lattices being generated by the RBC collaboration\cite{cdawson}.  

\section*{Acknowledgments}
I warmly thank the MILC collaboration for use of
their lattices, and in particular Doug Toussaint for his help. The
MILC code was used as a basis for these computations.

\end{document}